\documentclass[aps,apl,twocolumn]{revtex4}
\usepackage{amsfonts}
\usepackage{amsmath}
\usepackage{amsbsy}
\usepackage{graphicx}

\renewcommand\vec[1]{\ensuremath\boldsymbol{#1}}
\newcommand\rvec[1]{\ensuremath\bar{\boldsymbol{#1}}}

\begin{document}

\title{Multiferroic oxides-based flash-memory and spin-field-effect transistor}

\author{Chenglong Jia and  Jamal Berakdar}

\affiliation{Institut f\"ur Physik, Martin-Luther Universit\"at Halle-Wittenberg, 06120 Halle (Saale), Germany}

\begin{abstract}
We propose a modified  spin-field-effect transistor fabricated in a two dimensional electron gas (2DEG) formed at the surface of multiferroic oxides with a transverse helical magnetic order. The topology of the oxide local magnetic moments  induces a resonant momentum-dependent effective spin-orbit interaction acting
 on 2DEG. We show that spin polarization dephasing  is  strongly suppressed which is crucial for functionality. The carrier spin precession phase depend linearly on the magnetic spiral helicity. The latter is electrically controllable by virtue of the magento-electric effect. We also suggest a flash-memory device based on this structure.
\end{abstract}

\maketitle

Spin-based electronics, or  spintronics has been a growing area of research with a number of
 promising applications \cite{awscha}. A widely discussed device  is the spin-field-effect transistor   (spin-FET) proposed by Datta and Das \cite{Datta-Das}.
 In this device, electrons with a definite spin orientation leaving a ferromagnetic source traverse  a semiconductor-based 2DEG with a gate-controlled Rashba \cite{Rashba} spin-orbit interaction (SOI)  and arrive at a ferromagnetic drain. The \emph{on} and \emph{off} states are distinguished by $\pi$ phase difference in spin precession motion (due to SOI). A crucial point in this device is the  wave-vector ($\vec{k}$)-dependence of the Rashba SOI: the momentum scattering reorients the direction of the precession axis resulting a random effective magnetic field. This leads to an average spin dephasing and limits the practical functionality of the Datta-Das FET to the (quasi) ballistic regime. For systems with a Dresselhaus \cite{Dresselhaus} SOI in addition to the Rashba SOI,  another type of spin FET is proposed in Ref.\cite{RSFET1,RSFET2} when these two types of SOIs are exactly balanced:  The relaxation of spins oriented along the [110] axis is totally suppressed;
 non-magnetic scatterer cannot induce a spin flip enhancing thus the spin-coherence time, as demonstrated recently \cite{awscha2}. The disadvantage of such a resonant spin-FET is obviously that the current during the off state is not zero but approximately one-half of the on-current which underlies the view \cite{Cahay} that the present versions of spin-FET are not likely to be competitive with their electronic counterparts.

\begin{figure}[b]
\includegraphics[width=6cm,angle=-90]{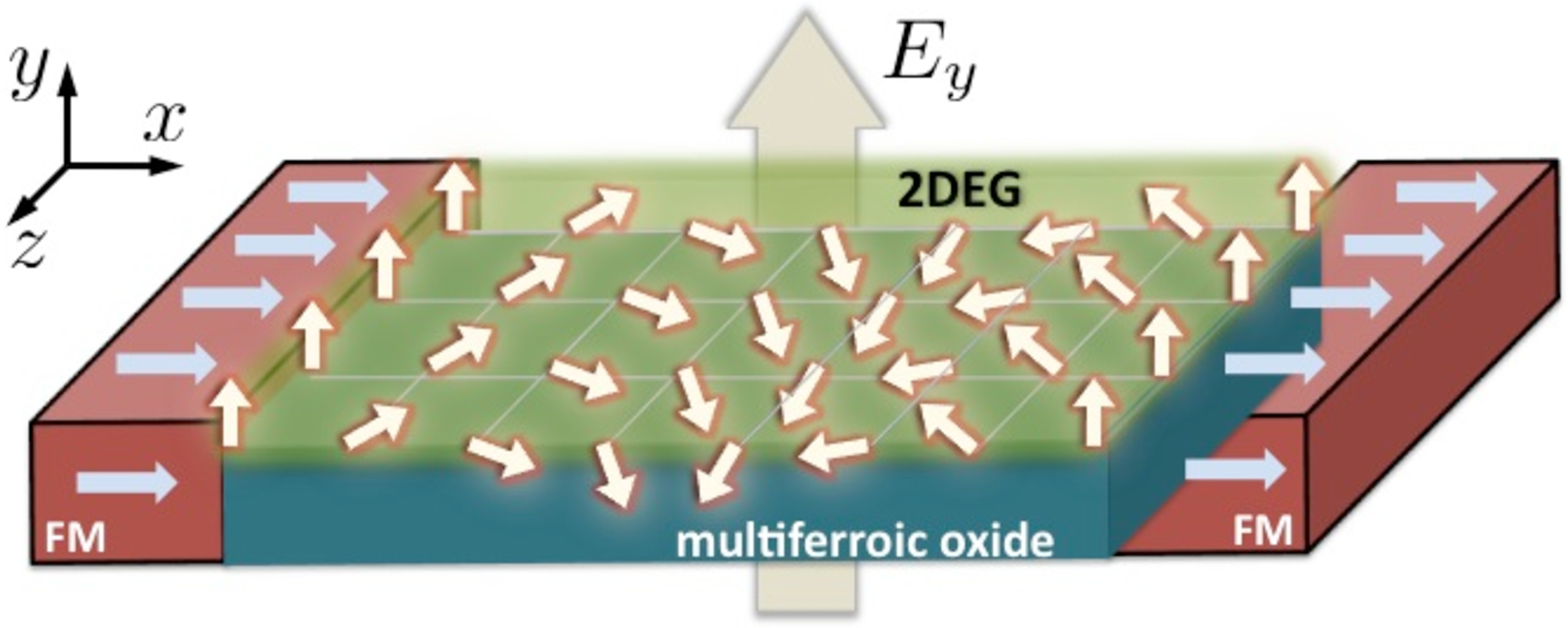}
\caption{(Color online) A schematics of proposed spin-FET device. Two ferromagnetic contacts are magnetized in the $x$-direction. The spiral plane of multiferroic oxide is perpendicular to the device, and the spin helicity is gate-controlled ($E_z$).}
\label{fig::SFET}
\end{figure}

Recently, 2DEG is experimentally realized at the interface of insulating  oxides \cite{Oxide0, Oxide1}.  Lateral  confinement and  patterning  \cite{Oxide2} allowed the demonstration of nanometer-sized tunnel junctions and field-effect transistors \cite{Oxide3}, which opens the way  for oxide-based Nanoelectronics \cite{Oxide2}.
 In this letter we show that utilizing  the 2DEG formed at the surface of  multiferroic oxides (Fig.\ref{fig::SFET}), the original Datta-Das device is operational even in the non-ballistic regime. Due to the topological structure of the local magnetic moment at the multiferroic interface (Fig.\ref{fig::SFET}), a traveling electron experiences an effective SOI that linearly depends on the carriers wave vector and on the helicity of the oxide's magnetic order \cite{ME-Oxide}. Such an effective SOI is in a complete analogy to the semiconductor case where    the Rashba SOI and Dresselhaus SOI have equal strengths. Therefore,  no decay of spin polarization coherence occurs during a momentum-dependent scattering. On the other hand, as shown experimentally, the helicity associated with the spin spiral structure of multiferroics is electrically \cite{E-Control} and/or magnetically \cite{H-Control} controllable based on the magneto-electric coupling \cite{ME}, and so is the resulting phase difference in spin precessional motion when traversing the 2DEG.

\begin{figure}[t]
\includegraphics[width=6cm]{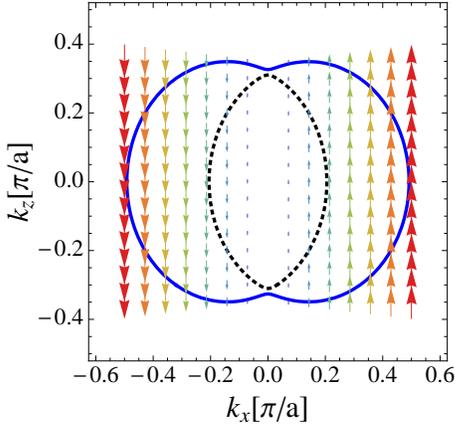}
\caption{(Color online) The arrows show the momentum-dependent magnetic fields induced by the effective spin-orbit interaction, $qk_x \sigma_z$. Also shown are the energy dispersions $E_{+}(\vec{k})$ (dotted curve) and $E_{-}(\vec{k})$ (solid curve) with $E= \frac{\hbar^2}{2ma^2}$, $J/E= 0.05$. The spiral wave vector is $q = 2\pi/7a$.  }
\label{fig::SOI}
\end{figure}

The proposal is sketched in Fig.\ref{fig::SFET}, a 2DEG is realized at the surfaces of a spiral multiferroic oxide such as the $ab$ plane of TbMnO$_3$ \cite{RMnO3}.
The local magnetic moments at the multiferroic surface reads $\vec{M}_r = M_0 \vec{n}_r$, where
$\vec{n}_r =  \left( \sin (\vec{q}_m \cdot \vec{r}),  \cos (\vec{q}_m \cdot \vec{r}), 0 \right)$
with $\vec{q}_m = (q,0,0)$ being the spin-wave vector of the spiral. The oxide local spin dynamics is much slower than that of the carriers in the 2DEG, so  one may assume at low-temperatures that the oxide local magnetic moments are classical and static. A conducting carrier is subject to an effective internal magnetic field generated by the magnetic spiral in the embedding medium \cite{swan,wannier,luttinger}. This results in a nonlocal vector potential. However, the energy contribution from the vector potential is much smaller than any relevant energy scale in the system and can be neglected \cite{ME-Oxide}. Thus, for the system Hamiltonian we employ the form
\begin{equation}
H=\frac{1}{2m}\vec{P}^2 + J \vec{n}_r \cdot \vec{\sigma}
\end{equation}
where $m$ is the effective electron mass, $J\vec{n}_r$ is the exchange field \cite{ME,exchange}, $J$ is the coupling strength, $\mathbf P$ is the momentum operator, and $\vec{\sigma}$ is the vector of the Pauli matrices.

Using the local gauge transformation in the spin space $U_g = e^{i \theta_r \sigma_z /2}$ one achieves  $U_g^{\dag} (J\vec{n}_r \cdot \vec{\sigma}) U_g = J\tilde{\sigma}_y$ (transformed quantities are labeled  by a tilde). This corresponds to a rotation of the local quantization axis as to align with $\vec{n}_r$ at each site. Note, $\sigma_z = \tilde{\sigma}_z$ because $[U_g, \sigma_z]=0$. As a result of the gauge transformation, an additional gauge field $\vec{A}_g = -i \hbar U_g^{\dagger} \nabla_{\vec{r}}  U_g = \hbar\tilde{\sigma}_z \vec{q}_m/2$ is introduced in the transformed kinetic energy \cite{gauge}.
The gauge field $\vec{A}_g $ depends only
on the geometry of the local magnetization at the oxide \cite{ME-Oxide}. Since  $\vec{A}_g \propto \vec{q}_m $
it can be changed electrically because $\vec{q}_m$ is gate-tunable, e.g. in the way sketched in fig.\ref{fig::SFET}.
The transformed single-particle Hamiltonian  of the 2DEG reads
\begin{eqnarray}
\tilde{H} = \frac{1}{2m}(\vec{P} + \vec{A}_g)^2 + J\tilde{\sigma}_y.
\end{eqnarray}
The parameters entering this Hamiltonian have the following realistic values for a 2DEG at oxides interfaces:
the lattice constant $a = 5 {\AA}$ and the  effective mass $m/m_e =10$  with $m_e$ being the free-electron mass \cite{Oxide1}. Removing a uniform energy displacement $\Delta E = \hbar^2 q^2/8m$, we rewrite the Hamiltonian in the form
\begin{equation}
\tilde{H} = \frac{\hbar^2}{2m}(k_x^2 +k_z^2 + q k_x \tilde{\sigma}_z)+ J \tilde{\sigma}_y
\label{Hg}
\end{equation}
From this  relation we infer that the influence of the spiral multiferroic interface on the 2DEG is subsumed
 in an effective SOI that depends linearly on $\vec{q}_m$ and $\vec{k}$
    (for $q \rightarrow 0$, i.e. for collinear spin case this SOI diminishes).
The  $\vec{k}$-dependence is analogous  to semiconductor-based 2DEG with the Rashba and Dresselhause SOI \cite{RSFET1,RSFET2} being equal, in which case  the effective magnetic field is oriented along the [110] axis irrespective of $\vec{k}$ and the  spin  traverses the [110] channel without flipping. In our multiferroic oxide system, the helical spin order lead to a zero average spin relaxation in $xy$-plane, the electron spin relaxation will be associated only with the electron diffusion along the $z$ direction \cite{s-relaxation}. As shown in Fig.\ref{fig::SOI}, the  resonant effective SOI  strongly suppresses the spin dephasing along the $z$-axis.

\begin{figure}[b]
\includegraphics[width=6cm,angle=-90]{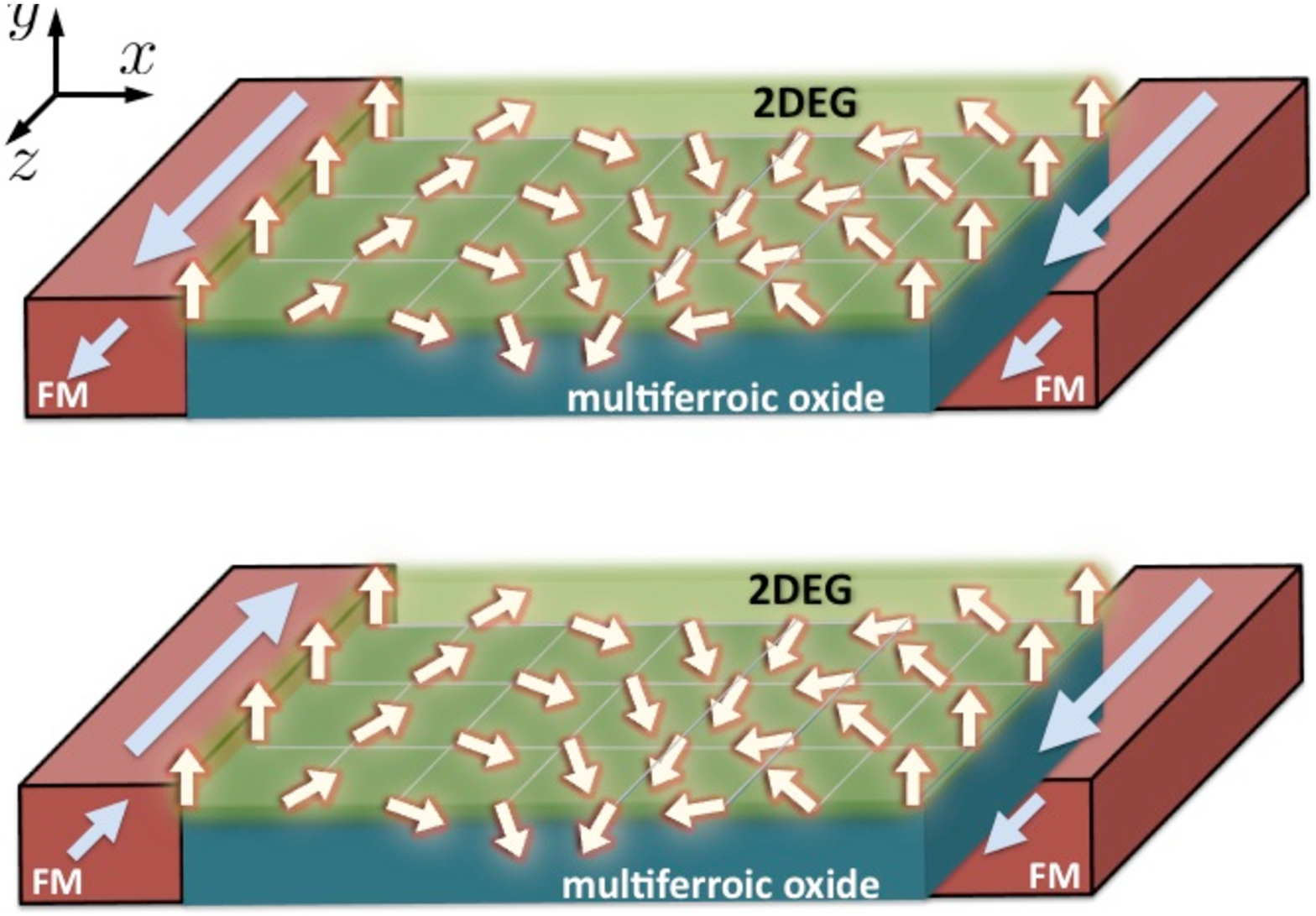}
\caption{(Color online) A proposal for a flash memory device. Different memory states are stored as the relative orientations of the ferromagnetic source and drain.}
\label{fig::memory}
\end{figure}

The eigenstates of $\tilde{H}$ read
\begin{eqnarray}
\label{eigenstates}
|\psi_{+} \rangle = e^{i \vec{k} \cdot  \vec{r}} \left( \begin{array}{cc}  \cos \frac{\phi}{2} \\ i \sin \frac{\phi}{2} \end{array} \right),\;
|\psi_{-} \rangle = e^{i \vec{k} \cdot  \vec{r}}  \left( \begin{array}{cc}  i \sin \frac{\phi}{2} \\ \cos \frac{\phi}{2} \end{array} \right)
\label{states}
\end{eqnarray}
where
\begin{eqnarray}
\cot \phi=  \frac{\hbar^2 q k_x}{2mJ},\; \; \cos \phi = \frac{qk_x  }{\sqrt{(2mJ/\hbar^2)^2 + (qk_x)^2}}.
\end{eqnarray}
The eigenenergies are
\begin{equation}
E_{\pm}(\vec{k}) = \frac{\hbar^2 \vec{k}^2}{2m} \pm \frac{\hbar^2 qk_x}{2m} \sqrt{1+ (\frac{2mJ}{\hbar^2 qk_x})^2 }.
\label{energies}
\end{equation}
Because of the effective spin-orbit coupling, the dispersions (\ref{energies}) are not parabolic but anisotropic. The  $\hat{x}$ and $\hat{z}$ are  the symmetry axes (cf. Fig.\ref{fig::SOI}). Although the spin states  Eq.(\ref{states}) are not independent of $\rvec{k}$, we still have no decay of  the spin-polarization coherence along $\hat{z}$ (in absence of  magnetic  scattering) just as in the case without magnetic field in Ref.\cite{RSFET1,RSFET2}. More importantly, as  evidenced by spin-polarized neutron scattering experiments \cite{E-Control},  one can change the helicity of the spiral magnetic order by applying  a small ($\sim 1 kV/cm$) transverse electric field. By doing so  we achieve  a gate-tunable phase difference $\Delta \theta$ in spin procession when the electron traverses the 2DEG, where
\begin{equation}
\Delta \theta = (k_{x}^{+}-k_{x}^{-}) L = q L.
\end{equation}
Here $L$ is the length of SOI active region, the double sign $\pm$ corresponds to the spin-up and spin-down branches. Note, in the above relation, we take an approximation of relatively weak exchange interaction in a high electron mobility transistor  (\emph{i.e.}, $J \ll \hbar^2 qk_x /2m$). In multiferroic oxide, a period of the spiral spin modulation is several lattice constant $a$, so $L$ is on the  nanometer scale.

Because the effective SOI always satisfies the resonance condition in our case
 another device, a flash memory is realizable as shown in Fig.\ref{fig::memory} in a similar
  manner as  proposal in Ref.\cite{RSFET2}.
  The "0" and "1" states are indicated by the relative orientation of the magnetization of the ferromagnetic source and drain.

  In summary, a multiferroic oxide-based 2DEG can be utilized as a nanometer-scale,
  decoherence-suppressed  spin field-effect transistor and as a nanometer flash-memory device.

\end{document}